\documentstyle[12pt]{ioplppt}

\begin{document}

\sloppy

\jl{2}

\letter{Multiphoton detachment from negative ions: new theory vs
experiment}

\author{G. F. Gribakin\ftnote{1}{E-mail: gribakin@newt.phys.unsw.edu.au}
and M. Yu. Kuchiev\ftnote{2}{E-mail: kuchiev@newt.phys.unsw.edu.au}}

\address{School of Physics, University of New South Wales,
Sydney 2052, Australia}

\begin{abstract}
In this paper we compare the results of our adiabatic theory
(Gribakin and Kuchiev, {\it Phys. Rev. A}, submitted for publication)
with other theoretical and experimental results, mostly for halogen
negative ions. The theory is based on the Keldysh approach. It shows that
the multiphoton detachment rates and the corresponding $n$-photon
detachment cross sections depend only on the asymptotic parameters $A$ and
$\kappa $ of the bound state radial wave function
$R(r)\simeq Ar^{-1}e^{-\kappa r}$. The dependence on $\kappa $ is very
strong. This is the main reason for the disagreement with some previous
calculations, which employ bound state wave functions with incorrect
asymptotic $\kappa $ values. In a number of cases our theoretical results
produces best agreement with absolute and relative experimental data.

\end{abstract}


\maketitle

In this paper we use the theory of multiphoton detachment from negative
ions developed in our previous work (Gribakin and Kuchiev 1996) to
calculate $n$-photon detachment cross sections and excess photon detachment
(EPD) spectra for some negative ions and frequencies, which have been
explored experimentally.

It has been shown that for a linearly polarized light the differential
$n$-photon detachment cross section for the electron in the initial state
$lm$ is
\begin{eqnarray}\label{sign}
\frac{\d \sigma ^{(lm)}_n}{\d\Omega }&=&\frac{pA^2\omega (2l+1)}
{4\pi ^2\sqrt{2n\omega }}\,
\frac{(l-|m|)!}{(l+|m|)!}\left| P_l^{|m|}
\left( \sqrt{1+p_\perp ^2/\kappa ^2}\right) \right| ^2
\left( \frac{\pi e}{nc\omega ^2}\right) ^n\nonumber \\
&\times &\frac{\exp (p_\parallel ^2/\omega )}{\sqrt{\kappa ^2+p_\perp ^2}}
\left[ 1+(-1)^{n+l+m}\cos \Xi \right] ~,
\end{eqnarray}
where $\kappa $ is determined by the initial energy of the electron
$E_0\equiv -\kappa ^2/2$, $p=\sqrt{2n\omega -\kappa ^2}$ is the
photoelectron momentum, $p_\parallel =p\cos \theta $ and
$p_\perp =p\sin \theta $ are its components parallel and perpendicular to
the field, $c\approx 137$ is the speed of light, $e=2.71\dots $,
$P_l^{|m|}$ is the associated Legendre function, and
\begin{equation}\label{Xi}
\Xi=(2n+1)\tan ^{-1}
\frac{p_\parallel }{\sqrt{\kappa ^2+p_\perp ^2}}+
\frac{p_\parallel \sqrt{\kappa ^2+p_\perp ^2}}{\omega }~.
\end{equation}
is the phase which determines the oscillatory pattern of the photoelectron
angular distribution. This effect is due to interference between the two
contributions to the photodetachment amplitude produced at the moments
of time when the field strength reaches its maximum (there are two such
instants in every period of the field).
The phase $\Xi $ varies with the ejection angle of the photoelectron 
in between $-\Xi_0$ and $\Xi_0$, where
$\Xi _0=(2n+1)\tan ^{-1} (p/\kappa )+p\kappa /\omega $, 
and can be quite large, even for the lowest-$n$ process,
$p\sim \sqrt{\omega }$, $\Xi _0\sim \sqrt{n}$.
Note that in accordance with the general symmetry properties, the cross
section is zero at $\theta =\pi /2$, when $n+l+m$ is odd.

After the electron detachment from a closed-shell negative ion the neutral
atom is left in either of the two fine-structure states with the total
angular momentum $j=l\pm \frac{1}{2}$, e.g., $^2P_{3/2}$ and $^2P_{1/2}$
for halogens. In this case the $n$-photon detachment cross
section summed over the projections of the angular momentum is given by
\begin{equation}\label{wj}
\frac{\d\sigma _n^{(j)}}{\d\Omega }=\frac{2j+1}{2l+1}\sum _{m=-l}^{l}
\frac{\d\sigma ^{(lm)}_n}{\d\Omega}~,
\end{equation}
where different values of $\kappa $ and binding energies $|E_0|$ should
be used for $j=l\pm \frac{1}{2}$, since the two sublevels have different
detachment thresholds. The main contribution to the sum in equation
(\ref{wj}) comes from $m=0$, since these orbitals are extended along the
direction of the field.

The sum over $m$ can be carried out analytically, 
taking into account the fact that $P_l^{|m|}$ in equation
(\ref{sign}) is a function of the imaginary
angle $\vartheta $, $\cos \vartheta =\sqrt{1+p_\perp ^2/\kappa ^2}$,
so that $\left[ P_l^{|m|}(\cos \vartheta )\right] ^*=(-1)^m
P_l^{|m|}(\cos \vartheta )$. The result is
\begin{equation}\label{dsj}
\fl \frac{\d \sigma ^{(j)}_n}{\d\Omega }=\frac{pA^2\omega (2j+1)}
{4\pi ^2\sqrt{2n\omega }}\left( \frac{\pi e}{nc\omega ^2}\right) ^n
\frac{\exp (p_\parallel ^2/\omega )}{\sqrt{\kappa ^2+p_\perp ^2}}
\left[ P_l\left( 1+2p_\perp ^2/\kappa ^2\right) +
(-1)^{n+l}\cos \Xi \right] ~,
\end{equation}
where $P_l$ is the Legendre polynomial, $P_0(x)=1$, $P_1(x)=x$, etc.

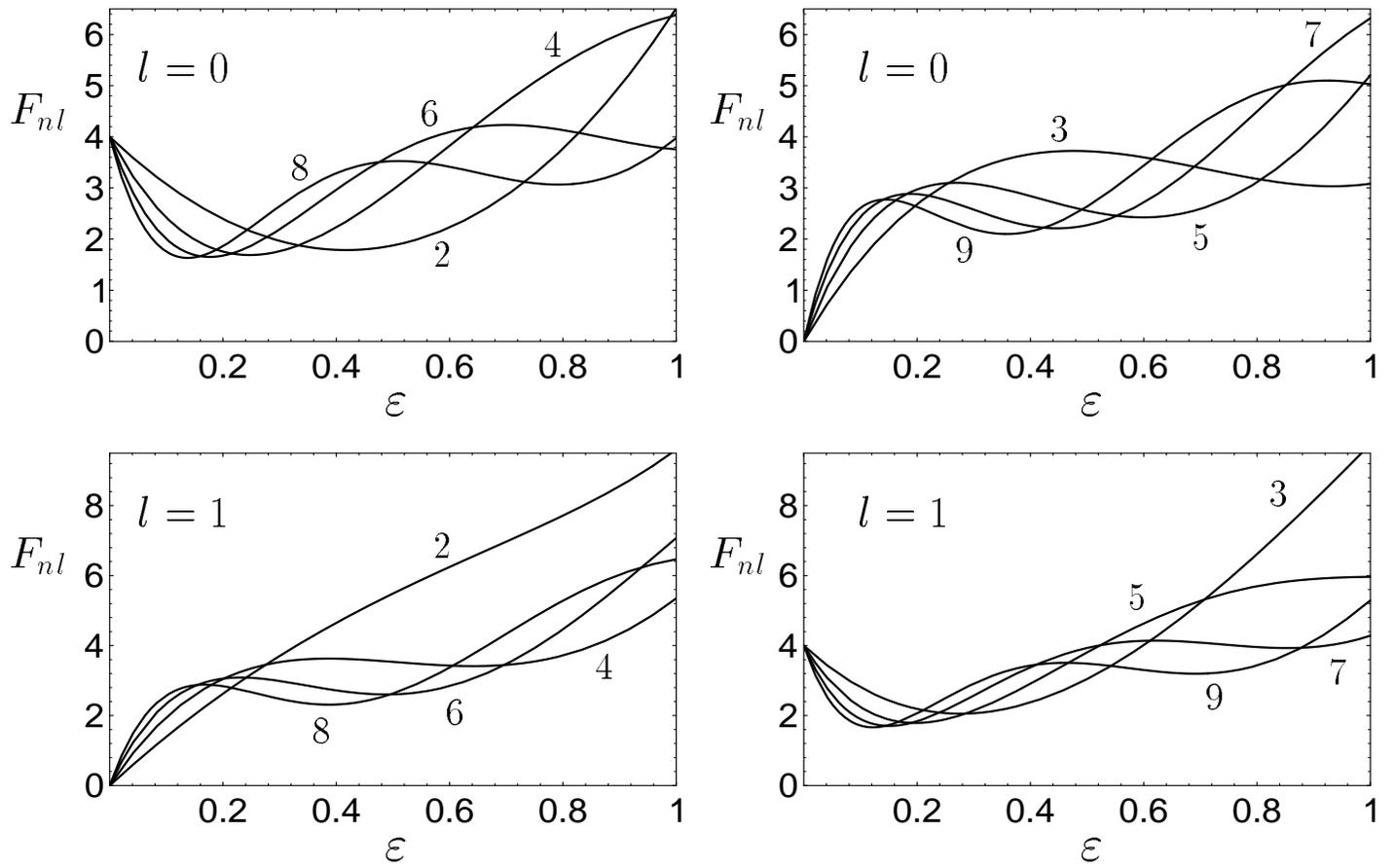
\begin{figure}
\caption{Dependence of the universal function $F_{nl}(\varepsilon )$
on the photoelectron energy in units of $\omega $ for the $n$-photon
detahcment of $s$ ($l=0$) and $p$ ($l=1$) electrons, for $n=2,\dots ,9$
(values of $n$ label curves in the plots).
The different behaviour of $F_{nl}(\varepsilon )$ at small
photoelectron energies $\varepsilon \ll 1$ ensures correct
threshold behaviour of the $n$-photon detachment cross section,
$\sigma _n\propto \varepsilon ^{1/2}$ for even $n+l$, and
$\sigma _n\propto \varepsilon ^{3/2}$ for odd $n+l$.\label{Ffunc}} 
\end{figure}

The total $n$-photon detachment cross section $\sigma _n^{(j)}$ is
obtained  by integrating equation (\ref{dsj}) over the emission angles
of the photoelectron. The result
can be presented in the following form
\begin{equation}\label{stot}
\sigma _n^{(j)}=(2j+1)\frac{pA^2}
{4\pi n}\left( \frac{\pi e}{nc\omega ^2}\right) ^n
F_{nl}(p^2/2\omega )~,
\end{equation}
where $F_{nl}$ is a universal function of the photoelectron energy in the
units of $\omega $,
\begin{eqnarray}\label{Fnl}
F_{nl}(\varepsilon )=\int _{-1}^{1}\frac{e^{2\varepsilon x^2}}
{\sqrt{1-\varepsilon x^2/n}}\left\{ P_l\left( 1+\frac{2\varepsilon (1-x^2)}
{n-\varepsilon}\right) +\right. \nonumber \\
\left.
+(-1)^{n+l}\cos \left[ (2n+1)\tan ^{-1}\left(\frac{x\sqrt{\varepsilon }}
{\sqrt{n-\varepsilon x^2}}\right) +
2x\sqrt{\varepsilon (n-\varepsilon x^2)}\right]\right\} \d x~.
\end{eqnarray}
Despite a somewhat cumbersome look, this dimensionless function is easy
to calculate, and we present it for $l=0,1$, $n=2$--9 in figure
\ref{Ffunc}. Note the different behaviour of $F_{nl}$ at small
$\varepsilon $, which provides a correct threshold law $\sigma \propto p$,
or $\sigma \propto p^3$, for even and odd $n+l$, respectively. This
depends on whether the lowest photoelectron partial wave is $s$ or $p$.
Equations (\ref{dsj}) and (\ref{stot}) and figure 1 predict a variety
of properties for the photodetachment from all negative ions.

The theory which has lead to equations (\ref{sign}) and (\ref{dsj}) shows
that the detachment process takes place when the electron is far from the
atomic core, at distances estimated as $r\sim \kappa ^{-1}\sqrt{2n}$
(Gribakin and Kuchiev 1996). This is why the probabilities can be expressed
in terms of the asymptotic parameters $A$ and $\kappa $ of the
bound-state wave function. All corrections due to electron-atom scattering
or electron correlations, e.g., the electron-atom polarizational
interaction, are expected to be suppressed by some powers of
$n$.

Equation (\ref{stot}) enables one to make simple estimates of saturation
intensities $I_{\rm S}$ defined as
$\sigma ^{(j)}_n(I_{\rm S}/\omega )^n\tau =1$,
where $\tau $ is the laser pulse duration,
\begin{equation}\label{Is}
I_{\rm S}=\frac{nc\omega ^3}{\pi e}
\left[ \frac{4\pi n}{(2j+1)pA^2F_{nl}\tau } \right] ^{1/n}.
\end{equation}
For low photoelectron energies $p^2/2\sim \omega $, $F_{nl}\sim 1$, the
$n$-dependence of $I_{\rm S}$ is basically determined by the first factor
with $\omega \approx |E_0|/n$,
\begin{equation}\label{In}
I_{\rm S}\propto n^{-2}.
\end{equation}
This dependence was first noticed in the numerical calculations by 
Crance (1988).

In what follows we use equations (\ref{dsj}) and (\ref{stot}) to calculate
the cross section for those negative ions and photon frequencies where
some experimental data are available. Parameters of the electron bound
state wave functions in these negative ions are listed in table
\ref{param}. Note that if after the detachment the neutral atom is left in
its ground state, then $\kappa $ is determined by the corresponding
electron affinity $E_{\rm a}$ (Hotop and Lineberger 1985),
$\kappa =\sqrt{2E_{\rm a}}$, and if the atom in the final state is excited
(e.g., to the upper fine-structure level), then
$\kappa =\sqrt{2(E_{\rm a}+\Delta )}$, where $\Delta $ is the energy of
the atomic excitation (Radtsig and Smirnov 1986). In both cases we are
dealing with the detachment of the valence electron, hence, the same value
of $A$ is used. In contrast with $\kappa $, the values of $A$ for most
ions are known only to within 10 \% or worse. In spite of the fact that the
present theory is strictly valid for large $n$, it will be applied
to $n$ as low as 2, where, we believe, it gives reasonable answers.

\begin{table}
\caption{Parameters of the negative ions used to calculate the $n$-photon
detachment cross sections.}\label{param}
\begin{indented}
\item[]\begin{tabular}{@{}lllllll}
\br
Ion & Term & Atom & $l$ & $j$ & $A$$^{\rm a}$ & $\kappa $ \\
\mr
O$^-$ & $^2P$ &  $^3P$ & 1 & 0.5$^{\rm b}$ & 0.65 & 0.328\\
Cu$^-$ & $^1S_0$ & $^2S_{1/2}$ & 0 & 0.5 & 1.2 & 0.301\\
Ag$^-$ & $^1S_0$ & $^2S_{1/2}$ & 0 & 0.5 & 1.3 & 0.3094\\
Au$^-$ & $^1S_0$ & $^2S_{1/2}$ & 0 & 0.5 & 1.3 & 0.4119\\
F$^-$ & $^1S_0$ & $^2P_{3/2}$ & 1 & 1.5 & 0.7 & 0.4998 \\
F$^-$ & $^1S_0$ & $^2P_{1/2}$ & 1 & 0.5 & 0.7 & 0.5035 \\
Cl$^-$ & $^1S_0$ & $^2P_{3/2}$ & 1 & 1.5 & 1.3 & 0.5156 \\
Cl$^-$ & $^1S_0$ & $^2P_{1/2}$ & 1 & 0.5 & 1.3 & 0.5233 \\
Br$^-$ & $^1S_0$ & $^2P_{3/2}$ & 1 & 1.5 & 1.4 & 0.4973 \\
Br$^-$ & $^1S_0$ & $^2P_{1/2}$ & 1 & 0.5 & 1.4 & 0.5300 \\
I$^-$ & $^1S_0$ & $^2P_{3/2}$ & 1 & 1.5 & 1.8 & 0.4742 \\
I$^-$ & $^1S_0$ & $^2P_{1/2}$ & 1 & 0.5 & 1.8 & 0.5423 \\
\br
\end{tabular}
\item[$^{\rm a}$] Values taken from Radtsig and Smirnov (1986) for
O$^-$, Cu$^-$, Ag$^-$, and Au$^-$, and from Smirnov and Nikitin (1988)
for F$^-$, Cl$^-$, Br$^-$, and I$^-$.
\item[$^{\rm b}$] This value gives correct cross sections for oxygen, when
the fine-structure is neglected.

\end{indented}
\end{table}

Figure 2(a) presents a comparison of the 2-photon detachment cross sections
for O$^-$, Cu$^-$, Ag$^-$ and Au$^-$ measured by Stapelfeldt \etal
(1991a, 1991b) at $\omega =1.165$ eV, with our values, and the
theoretical result of Robinson and Geltman (1967) for O$^-$. In the latter
calculation the single-electron wave functions were calculated in a model
polarization potential chosen to reproduce the correct value of the
electron affinity. Figure 2(b) shows the 3-photon cross sections
for F$^-$, Br$^-$ and I$^-$, as measured by Blondel \etal (1989) and
Kwon \etal (1989) (F$^-$ only), together with the results of our
calculations and those of Crance (1988), where the Hartree-Fock (HF) wave
functions and plane waves were used to describe the initial and the final
state of the electron, respectively. Two features are most obvious.
First, our theoretical values are considerably higher than the
experimental ones, and second, there is a much better agreement between
the relative cross sections, as given by the theory and experiment.
There is also a reasonable agreement between the two calculations for
O$^-$, Br$^-$ and I$^-$.

The large disagreement between our value for F$^-$ and that of Crance 1988
is, as discussed in Gribakin and Kuchiev (1996),
due to the incorrect behaviour of the HF bound state wave
function, which falls off much faster than the true one (the HF
value of $\kappa $ for F$^-$ is 0.6, vs the true $\kappa =0.5$). Note that
among the halogen negative ions this discrepancy is the largest for F$^-$.
Accordingly, the theoretical cross sections for Br$^-$ and I$^-$ are in
better agreement (HF values of $\kappa =0.528,~0.508$
respectively; compare to the true values of $\kappa =0.497,~0.474$ for
$j=\frac{3}{2}$,
see table \ref{param}). It is worth pointing out that the calculations
for halogens that employed HF wave functions of the photoelectron
(Crance 1987a, 1987b) produced cross sections close to those obtained
within the plane-wave approximation. From the point of view of our theory
this is a consequence of the importance of large distances for the
multiphoton detachment. Therefore, we have to conclude that the agreement
that the theoretical and experimental values had enjoyed for F$^-$
was probably coincidental, 
caused by the significant  error in the previous calculations. 
On the other hand, the relative magnitudes of the cross
sections for F$^-$, Br$^-$ and I$^-$ given by our theory are very close
to the measured ones.

The minimal number of quanta needed for the detachment of Cl$^-$ at the
Nd:YAG laser frequency is four. In this instance the calculated
value $\sigma _4=10.2\times 10^{-124}$~cm$^8$s$^3$ is also greater
than the experimental
$\sigma _4=0.97^{+0.68}_{-0.41}\times 10^{-124}$~cm$^8$s$^3$
(Blondel and Trainham 1989), as well as that by Crance (1988),
$\sigma _4=5.6\times 10^{-124}$~cm$^8$s$^3$. In order to compare the
results for all four halogens at this frequency, one can calculate
the saturation intensities $I_{\rm S}$ for some $\tau $, e.g.,
$\tau =2\pi /E_{\rm a}$, as in Blondel and Trainham 1989. Using our
cross sections and the data from Blondel and Trainham 1989 we complete
table \ref{IS}.
\begin{table}
\caption{Saturation intensities for the multiphoton detachment of the
halogen negative ions by $\omega =1.165 $ eV linearly polarized light.}
\label{IS}
\begin{indented}
\item[]\begin{tabular}{@{}llllll}
\br
Ion & $n$ & $\tau $ & $I_{\rm S}^{\rm exp }$ & $I_{\rm S}^{\rm theor}$ &
$R=I_{\rm S}^{\rm exp }/I_{\rm S}^{\rm theor}$ \\
& & fs & $10^{13}$ W/cm$^2$ & $10^{13}$ W/cm$^2$ & \\
\mr
F$^-$ & 3 & 1.22 & 4.4 & 2.50 & 1.76 \\
Cl$^-$ & 4 & 1.14 & 3.2 & 1.79 & 1.79 \\
Br$^-$ & 3 & 1.23 & 3.2 & 1.72 & 1.86 \\
I$^-$ & 3 & 1.35 & 2.4 & 1.23 & 1.95 \\
\br
\end{tabular}
\end{indented}
\end{table}
The existing discrepancy between the theory and experiment is characterised
by the ratio $R$, which remains almost constant, $R\approx $1.8--1.9.

In a more recent work by Davidson \etal (1992) the 2- and 3-photon
detachment cross sections for Cl$^-$ by 2 eV photons have been measured.
Their value $\sigma _2=16^{+29}_{-8}\times 10^{-50}$ cm$^4$s is in
good agreement with ours, $\sigma _2=9.44\times 10^{-50}$ cm$^4$s, 
whereas the values predicted by other theories are notably smaller,
$\sigma _2=5.5\times 10^{-50}$ cm$^4$s (Crance 1987), and
$\sigma _2=2.5\times 10^{-50}$ cm$^4$s (Jiang and Starace 1987).
Our cross section also agrees with the result of a much earlier measurement
by Trainham \etal 1987 at a smaller photon energy, $\omega =1.874$ eV,
$\sigma _2=1.3\pm 0.9\times 10^{-50}$ cm$^4$s (experiment),
$\sigma _2=2.19\times 10^{-50}$ cm$^4$s (theory). The result of Robinson
and Geltman (1967) is $\sigma _2=1.68\times 10^{-50}$ cm$^4$s.
The 3-photon cross sections from Davidson \etal (1992),
$\sigma _3=1.84^{+6.3}_{-1.2}\times 10^{-82}$ cm$^6$s$^2$, is also
close to our calculation, $\sigma _3=10.35\times 10^{-82}$ cm$^6$s$^2$.

In the earlier experimental work devoted to the EPD Davidson \etal (1991)
measured the ratio between the
5-photon and 4-photon detachment signals in Cl$^-$ at $\omega =1.165$ eV.
If the energy of the laser pulse is below the saturation limit, then
for a pulse with a Gaussian spatial and temporal profile this ratio is
given by $\left( \frac{4}{5}\right) ^{3/2}(\sigma _5/\sigma _4)I/\omega $,
where $I$ the peak intensity in the pulse. Indeed, the experimental points
in figure 3 of Davidson \etal 1991 display an approximately linear
dependence on the intensity. The corresponding slope can be estimated
as 0.0415, if the intensity is in units of $10^{12}$ W/cm$^2$, which is
quite close to our theoretical estimate of 0.0466. When the cross sections
of Crance 1988 are used, a value of 0.053 is obtained. The difference
between our theory and that based on the HF description of the negative
ion is not as large for the ratio of the cross sections, as it is for
the cross sections themselves. At the peak intensity of
$2.4\times 10^{12}$~W/cm$^2$, well into the saturation regime, we
calculated the ratio of 4- and 6-photon detachments to be 120, with the
experimental value between 70 and 200.

The EPD was also measured for Au$^-$, where 2-, 3-, and possibly
4-photon detachment signals were observed at $\omega =1.165$~eV
(Stapelfeldt \etal 1991). However,
the interpretation of that experiment is not straightforward, because
already at $\frac{1}{15}$ of the maximal laser intensity of
$3\times 10^{12}$ W/cm$^2$ achieved in that experiment the $n=2$ channel is
closed due to the ponderomotive threshold shift.

Equation (\ref{dsj}) can be used to calculate angular distributions of
photoelectrons studied by Blondel \etal (1991,1992)
and Dulieu \etal (1995). For the 2-photon detachment from halogen negative
ion at $\omega =2.331$~eV the shapes of the differential cross sections
from equation (\ref{dsj}) are close to those produced by the
`HF $+$ plane-wave' calculations presented in figure 3 of
Blondel \etal 1992, the difference in the absolute values aside. They
reproduce well the features of the experimental data.
In some cases our calculations
are in better agreement with the experiment, e.g., for Cl$^-$,
Br$^-$ ($^2P_{3/2}$ final state), or F$^-$, see figure 3. Note that for
F$^-$ the analytical result (\ref{dsj}) is also in good agreement with
the numerical calcultions of Pan and Starace (1991), where
electron correlations in the initial, intermediate and final states of
the 2-photon process have been taken into account.
The largest discrepancy between the theory and experiment is observed for
I$^-$.  Similar conslusions can be drawn
from the comparison of our angular distributions at $\omega =1.165$~eV
(figure 3 of Gribakin and Kuchiev 1996) with the experimental results 
for F$^-$, $n=3\,,4\,,\,5$, Cl$^-$, $n=4$, Br$^-$, $n=3$,
and I$^-$, $n=3$ (Blondel \etal 1991, Dulieu \etal 1995).
Overall, the simple analytical expression (\ref{dsj})
reproduces remarkably  well the complicated oscillatory pattern of the
photoelectron angular distributions.

In summary, our theory of multiphoton detachment in a strong
laser field (Gribakin and Kuchiev 1996) yields very simple formulae
for the $n$-photon detachment cross sections and photoelectron angular
distributions. Apart from $\omega $ and $n$ they depend only on the
electron orbital angular momentum $l$ and the two constants $A$ and
$\kappa $ which characterize the behaviour of the bound-state wave
function at large distances. The correct asymptotic behaviour of the
bound-state wave fucntion at large distances is crucial for obtaining
correct absolute values of the $n$-photon cross sections. 
The fact that the theory is based on the properties of the wave
function at large separations from the atom make it very reliable.
At the same time the simple structure of our final results permits 
make quick estimates of the cross sections for  any negative ion.
We have applied the theory to the negative ions and processes
studied experimentally, and found that our results are in reasonable,
and in some cases, very good agreement with the experimental data.
Some discrepancies found call for more accurate
absolute experimental values. 
When more sophisticated calculations are performed,
our formulae can be used as a benchmark, to demonstrate the role of
various possible corrections,  which are expected be small for the
processes considered. Of course, there are other
processes where many-electron correlations  play a decisive role, e.g.,
multiple photodetachment or ionization (Kuchiev 1987, 1995, 1996).
		
\ack
We wish to acknowledge support of the Australian Research Council.

\section*{References}
\begin{harvard}
\item[] Blondel C and Trainham R 1989 {\it J. Opt. Soc. Am. B} {\bf 6} 1774

\item[] Blondel C, Champeau R-J, Crance M, Crubellier A, Delsart C and
Marinescu D 1989 {\it J. Phys. B: At. Mol. Opt. Phys.} {\bf 22} 1335

\item[] Blondel C, Crance M, Delsart C and Giraud A 1991 {\it J. Phys. B:
At. Mol. Opt. Phys.} {\bf 24} 3575

\item[] \dash 1992 {\it J. Physique II} {\bf 2} 839

\item[] Blondel C, Champeau R-J, Crance M, Crubellier A, Delsart C and
Marinescu D 1989 {\it J. Phys. B: At. Mol. Opt. Phys.} {\bf 22} 1335

\item[]Crance M 1987a {\it J. Phys. B: At. Mol. Phys.} {\bf 20} L411

\item[]\dash 1987b {\it J. Phys. B: At. Mol. Phys.} {\bf 20} 6553

\item[]\dash 1988 {\it J. Phys. B: At. Mol. Opt. Phys.} {\bf 21} 3559

\item[] Davidson M D, Muller H G and van Linden van den Heuvell H B
1991 {\it Phys. Rev. Lett.} {\bf 67} 1712

\item[] Davidson M D \etal 1992 {\it J. Phys. B: At. Mol. Opt. Phys.}
{\bf 25} 3093

\item[] Dulieu F, Blondel C and Delsart C 1995 {\it J. Phys. B:
At. Mol. Opt. Phys.} {\bf 28} 3861

\item[] Gribakin G F and Kuchiev M Yu 1996 {\it Phys. Rev. A}
submitted for publication

\item[] Hotop H and Lineberger W C 1985 {\it J. Phys. Chem Ref. Data}
{\bf 14} 731

\item[] Kuchiev M Yu 1987 {\it Pis'ma Zh. Eksp. Teor. Fiz.} {\bf 45}
319 [{\it JETP Lett.} {\bf 45} 404]

\item[]\dash 1995 {\it J. Phys. B: At. Mol. Opt. Phys.} {\bf 28} 5093

\item[]\dash 1996 {\it Phys. Lett. A} {\bf 212} 77

\item[] Kwon N, Armstrong P S, Olsson T, Trainham R and Larson D J 1989
{\it Phys. Rev. A} {\bf 40} 676

\item[] Landau L D and Lifshitz E M Quantum Mechanics

\item[] Nikitin E E and Smirnov B M 1988 {\it Atomic and Molecular
Processes} (Moscow: Nauka) p~283

\item[] Radtsig A A and Smirnov B M 1986 {\it Parameters of Atoms and
Atomic Ions} (Moscow: Energoatomizdat) 

\item[] Robinson E J ans Geltman S 1967 {\it Phys. Rev.} {\bf 153} 4

\item[] Stapelfeldt H, Brink C and Haugen H K 1991a {\it J. Phys. B:
At. Mol. Phys.} {\bf 24} L437

\item[] Stapelfeldt H, Balling P, Brink C and Haugen H K 1991b
{\it Phys. Rev. Lett.} {\bf 67} 1731

\item[] Trainham R, Fletcher G D and Larson D J 1987 {\it J. Phys. B:
At. Mol. Phys.} {\bf 20} L777

\end{harvard}

\Figures

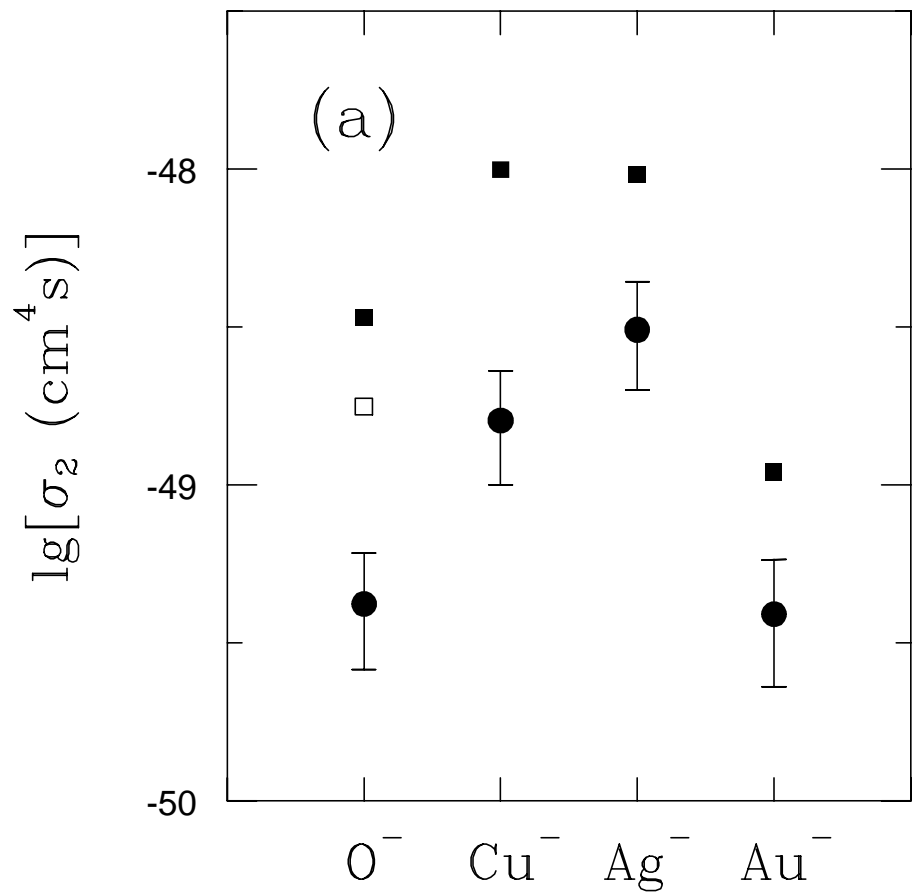
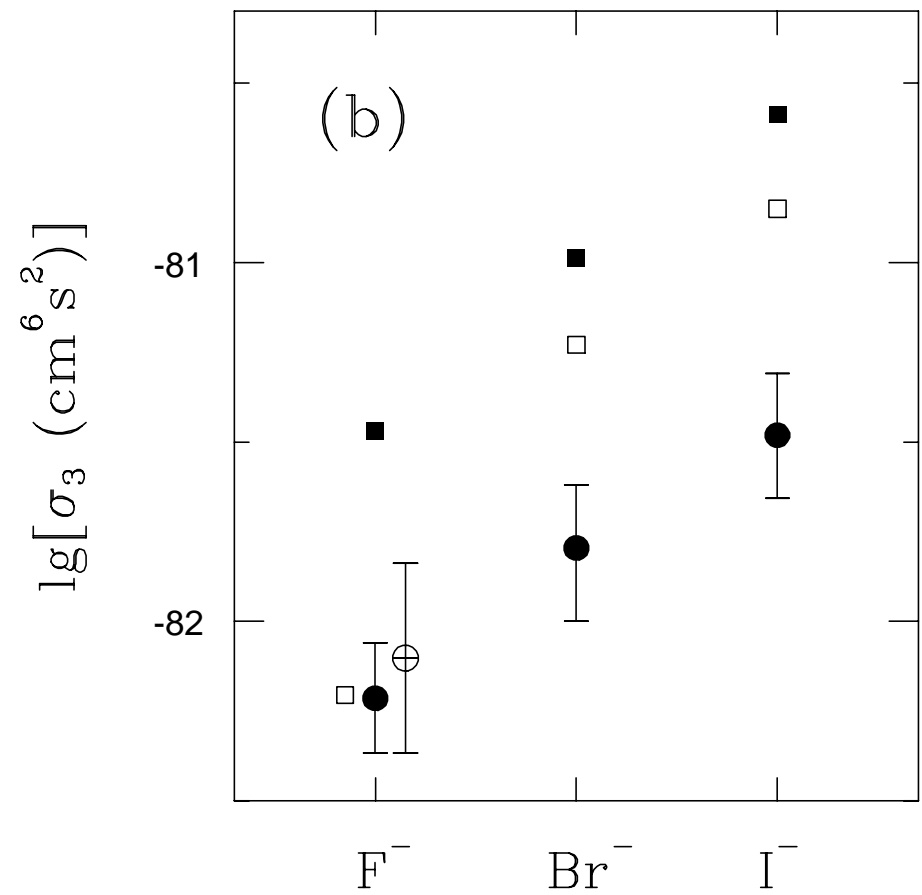
\begin{figure}
\caption{Comparison of the theoretical and experimental $n$-photon
detachment cross sections at $\omega =1.165$ eV. (a) $n=2$, \fullsqr ~, our
calculation; \opensqr ~, calculation by Robinson and Geltman
(1967), \protect \fullcirc ~, experiment by Stapelfeldt \etal (1991).
(b) $n=3$, \fullsqr ~, our calculation; \opensqr ~, calculation by Crance
(1988); \protect \fullcirc ~, experiment by Blondel \etal (1989);
\protect \opencirc ~, experiment by Kwon \etal (1989).\label{csecs}}
\end{figure}

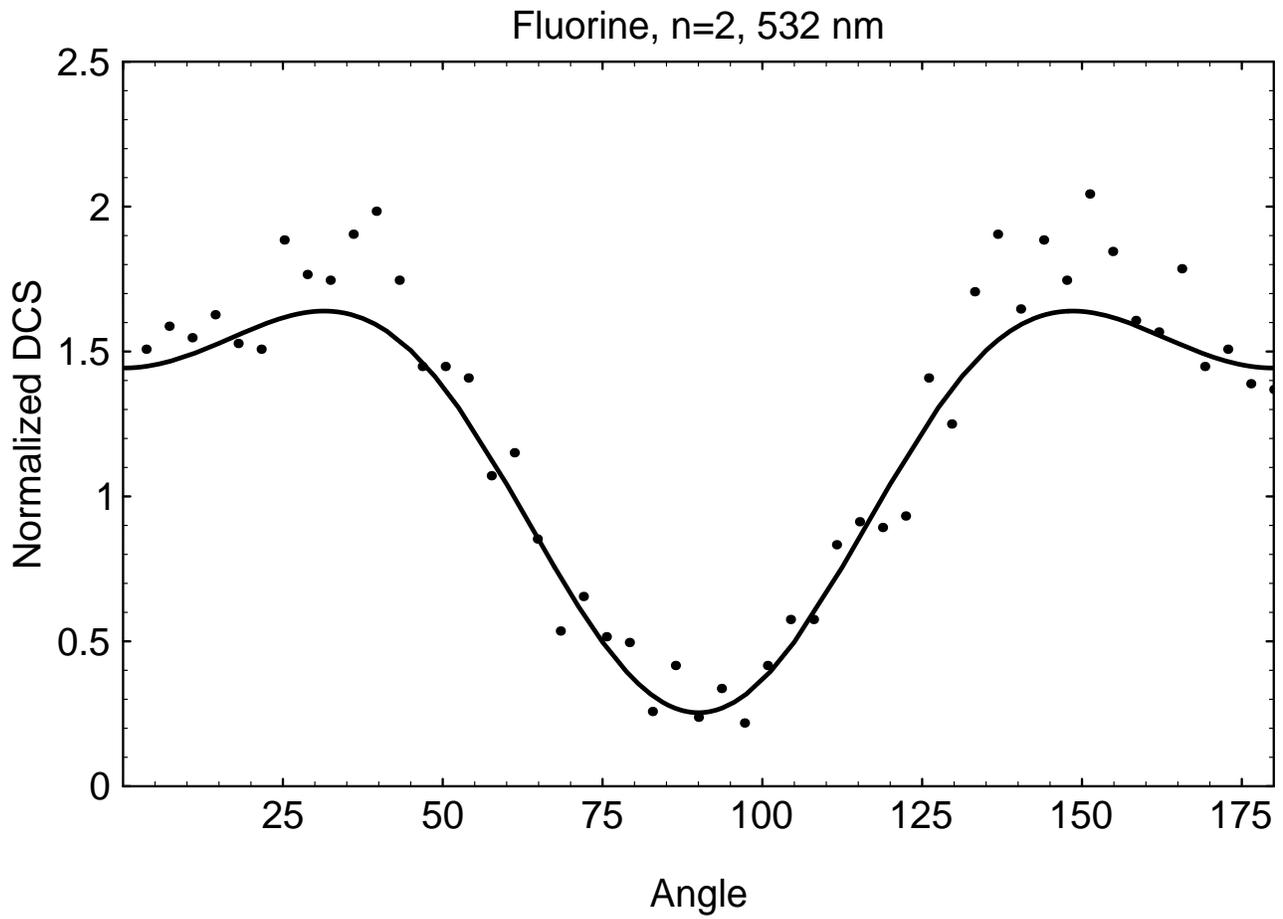
\begin{figure}
\caption{Normalized differential 2-photon detachment cross sections
$4\pi \sigma ^{-1}\d \sigma /\d \Omega $ for F$^-$ at $\omega =2.33$ eV.
Experimental results by Blondel \etal (dots) taken from Pan and Starace
(1991), and close to those given in Blondel \etal (1992); \full ~,
our calculation (sum of the $j=3/2$ and $j=1/2$ cross sections). 
\label{difsec}}
\end{figure}

\end{document}